\documentclass[prb,onecolumn,superscriptaddress,showpacs,floatfix]{revtex4}

\usepackage{epsfig,dcolumn,amsmath,latexsym}

\usepackage{subfigure}
\usepackage[normalem]{ulem}
\usepackage{cancel}
\usepackage{multirow}

\def\k233{K$_2$Cr$_3$As$_3$}

\begin{document}

\title{Electronic structure of quasi-one-dimensional superconductor \k233 from first-principles calculations} \preprint{1}

 \affiliation{Condensed Matter Group,
  Department of Physics, Hangzhou Normal University, Hangzhou 310036, P. R. China}
 \affiliation{Department of Physics, Zhejiang University, Hangzhou 310013, P. R. China}

\author{Hao Jiang}
 \affiliation{Department of Physics, Zhejiang University, Hangzhou 310013, P. R. China}
\author{Guanghan Cao}
 \affiliation{Department of Physics, Zhejiang University, Hangzhou 310013, P. R. China}
 \affiliation{Collaborative Innovation Centre of Advanced Microstructures, Nanjing University, Nanjing 210093, China}

\author{Chao Cao}
 \email[E-mail address: ]{ccao@hznu.edu.cn}
 \affiliation{Condensed Matter Group,
  Department of Physics, Hangzhou Normal University, Hangzhou 310036, P. R. China}
\date{Dec 3, 2014}

\begin{abstract}
The electronic structure of quasi-one-dimensional superconductor \k233 is studied through systematic first-principles calculations. The ground state of \k233 is paramagnetic. Close to the Fermi level, the Cr-3d$_{z^2}$, d$_{xy}$, and d$_{x^2-y^2}$ orbitals dominate the electronic states, and three bands cross $E_F$ to form one 3D Fermi surface sheet and two quasi-1D sheets. The electronic DOS at $E_F$ is less than 1/3 of the experimental value, indicating a large electron renormalization factor around $E_F$. Despite of the relatively small atomic numbers, the antisymmetric spin-orbit coupling splitting is sizable ($\approx$ 60 meV) on the 3D Fermi surface sheet as well as on one of the quasi-1D sheets. Finally, the imaginary part of bare electron susceptibility shows large peaks at $\Gamma$, suggesting the presence of large ferromagnetic spin fluctuation in the compound.
\end{abstract}

\maketitle

\section*{Introduction}

The effect of reduced dimensionality on the electronic structure, especially on the ordered phases, has been one of the key issues in the condensed matter physics. For a  one-dimension (1D) system with short-range interactions, the thermal and quantum fluctuations prevent the development of a long-range order at finite temperatures\cite{PhysRev.158.383,PhysRevLett.17.1133}, unless finite transverse coupling is present, i.e. the system becomes quasi-1D\cite{PhysRevB.27.5856}. The scenario is exemplified with the discovery of the quasi-1D superconductor including the Bechgaard salts (TMTSF)$_2X$\cite{TMTSF_1}, $M_2$Mo$_6$Se$_6$ ($M$=Tl, In)\cite{Brusetti1988181}, Li$_{0.9}$Mo$_6$O$_{17}$\cite{Greenblatt1984671,PhysRevLett.108.187003}, etc. In general, the Bechgaard salts exhibits charge-density-wave or spin-density-wave instabilities and superconductivity emerges once the instability is suppressed by applying external pressure. For the molybdenum compounds, although the density-wave instabilities were absent in the superconducting samples, they were observed in the closely related $A_2$Mo$_6$Se$_6$ and $A_{0.9}$Mo$_6$O$_{17}$ ($A$=K, Rb, etc). Therefore, a "universal" phase diagram\cite{Wilhelm:2001uq,TMT_review} suggests the superconductivity is in close proximity to the density-wave and most probably originated from non-phonon pairing mechanism. However, the determination of the pairing symmetry for these compound was difficult due to the competition between several energetically close ordering parameters\cite{PhysRevB.79.224501}.

  \begin{figure}
    \subfigure[3D view]{
      \includegraphics[width=5cm]{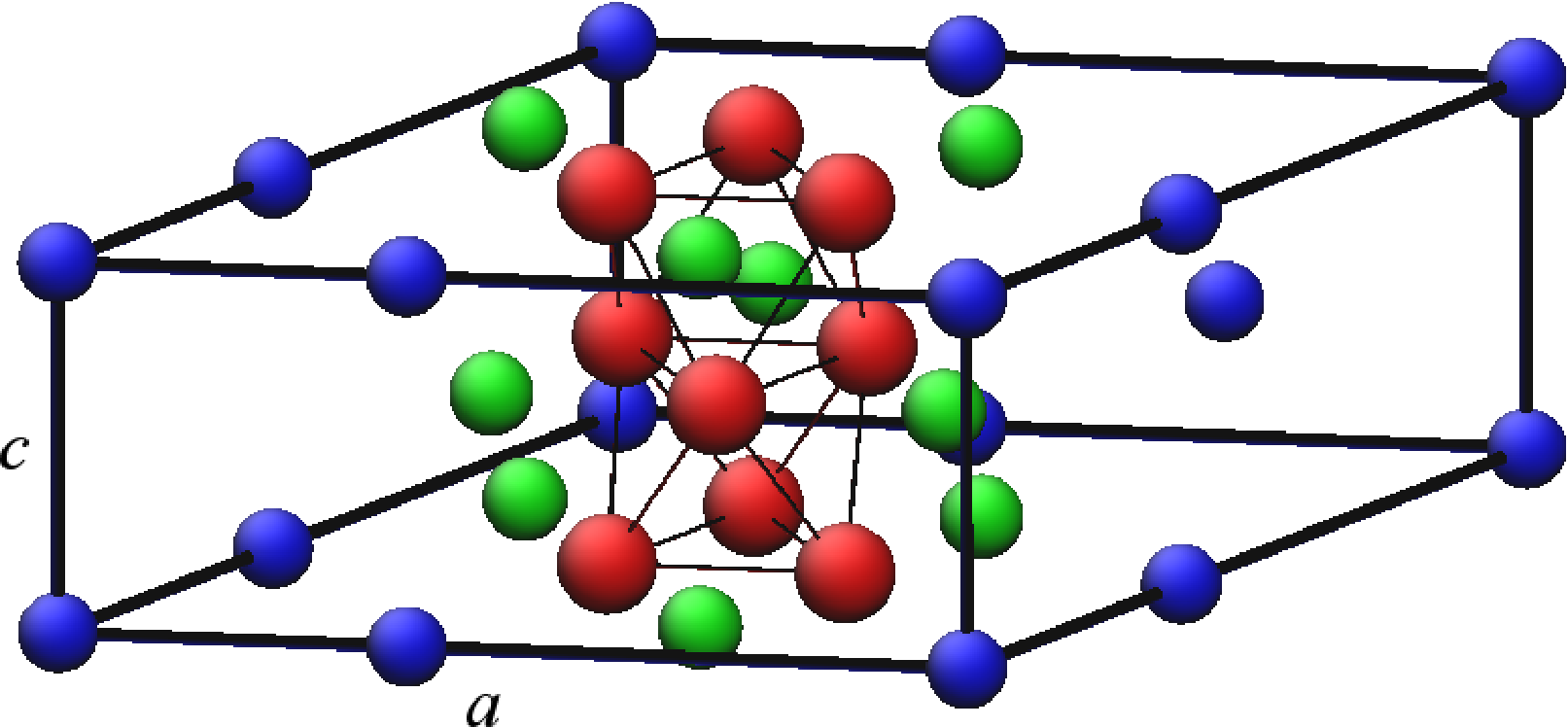}
      \label{fig:geometry}}
    \subfigure[Top view]{
      \includegraphics[width=3cm]{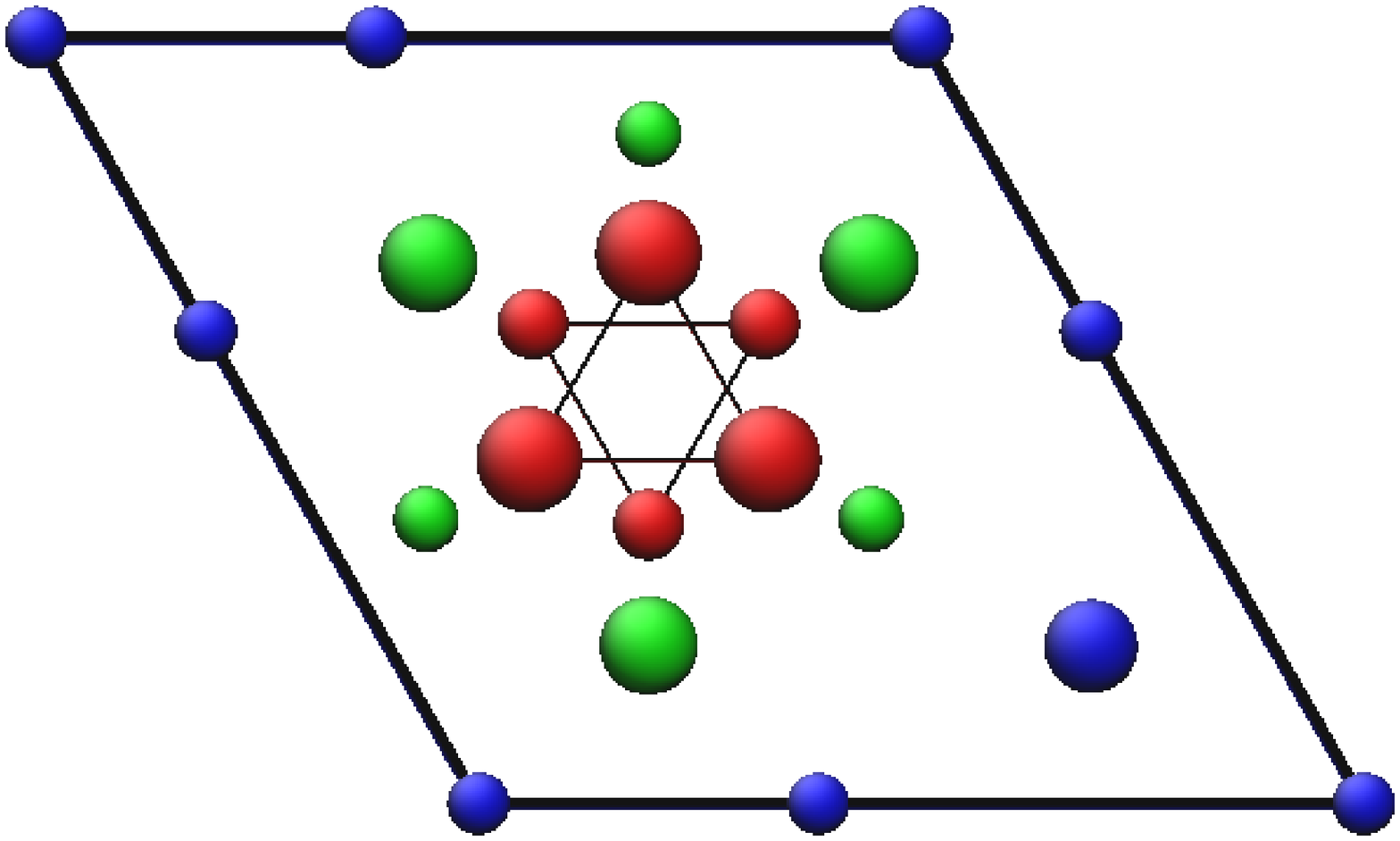}}
    \subfigure[IAF]{
      \includegraphics[width=1.5cm]{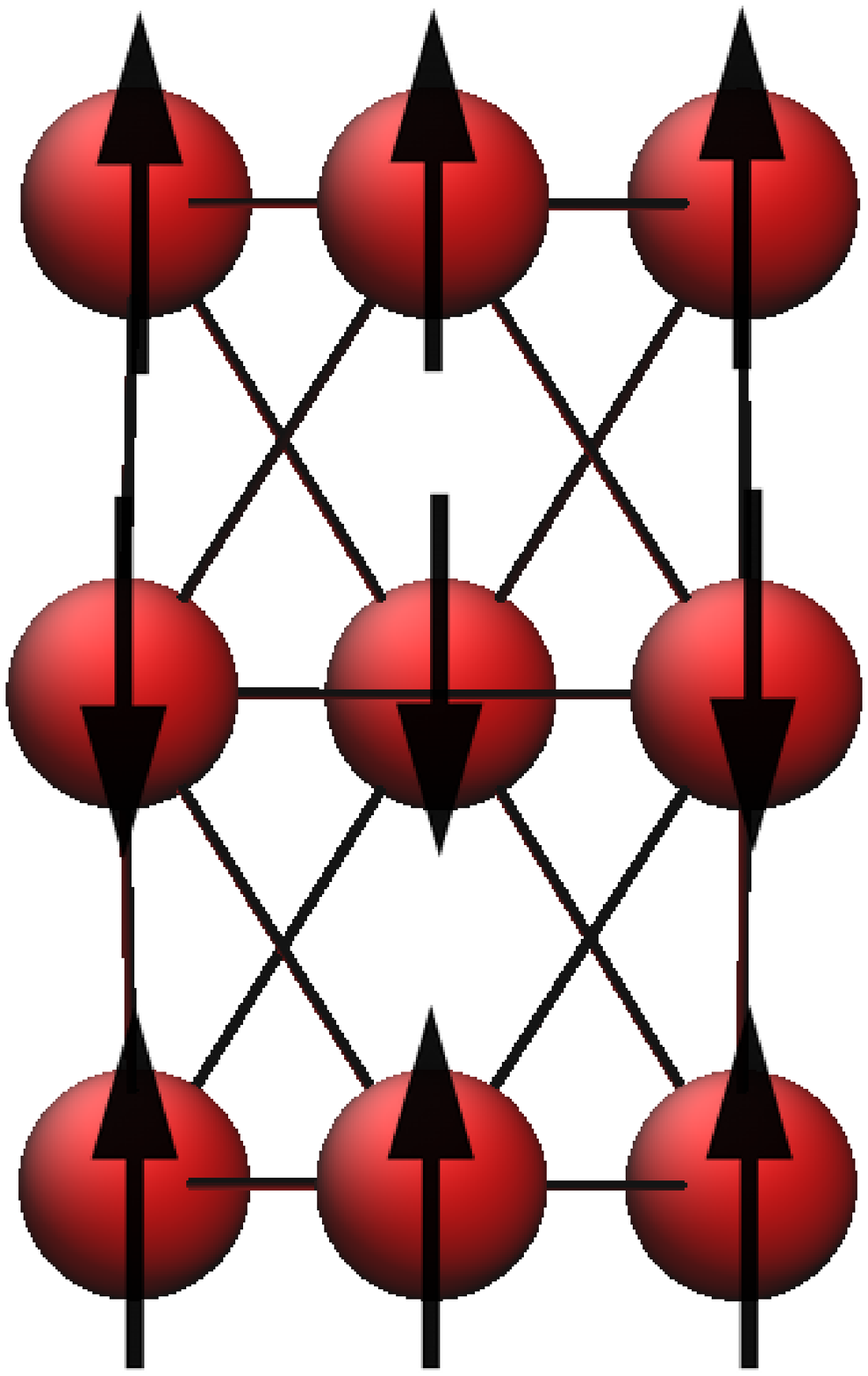}
      \label{fig:magiaf}}
    \subfigure[NCL]{
      \includegraphics[width=2cm]{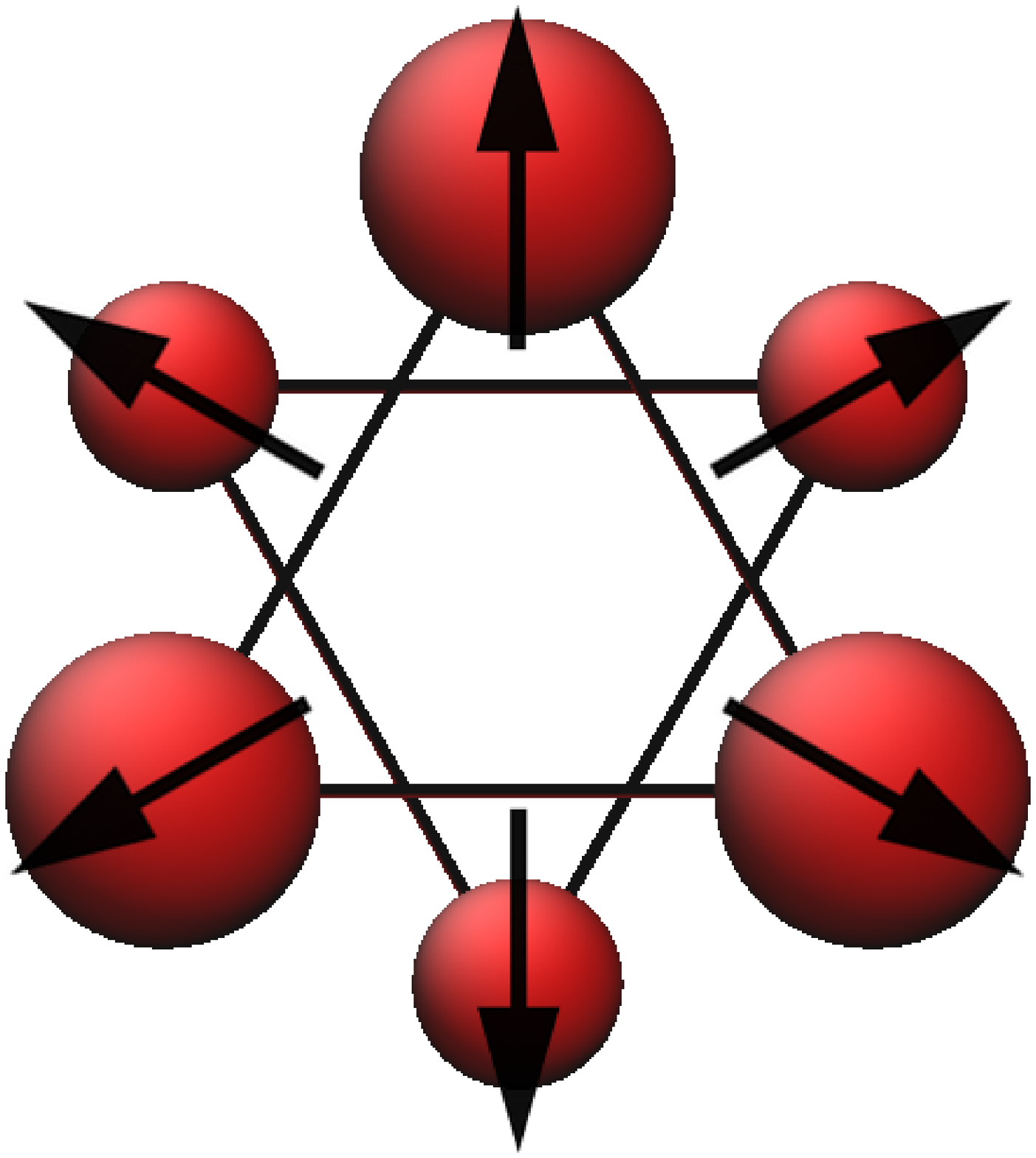}
      \label{fig:magncl}}
    \subfigure[IOP]{
      \includegraphics[width=2cm]{fig1e}
      \label{fig:magiop}}
    \subfigure[CLK]{
      \includegraphics[width=1.9cm]{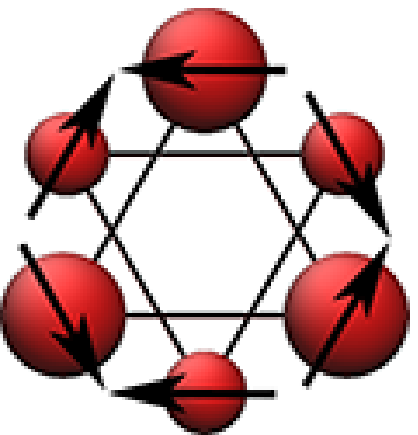}
      \label{fig:magclk}}
    \caption{(a-b) Crystal structure of \k233\ and (c-f) the considered magnetic states. The blue atoms are K, red atoms Cr, green atoms As. In panel (b), the larger spheres denote atoms at $z$=0.5 (Cr$^{\mathrm{II}}$, As$^{\mathrm{II}}$ and K$^{\mathrm{II}}$) and the smaller ones are at $z$=0.0 (Cr$^{\mathrm{I}}$, As$^{\mathrm{I}}$ and K$^{\mathrm{I}}$). In panels (c) and (d), only Cr atoms are shown for a clear view. IAF, NCL, IOP and CLK refer to interlayer antiferromagnetism, noncollinear antiferromagnetism, in-out state antiferromagnetism and chiral-like antiferromagnetism, respectively. \label{fig:geomag}}
  \end{figure}

Recently, a new family of superconducting compounds $A_2$Cr$_3$As$_3$ ($A$=K, Rb, Cs) was reported\cite{arxiv:1412.0067,PhysRevB.91.020506,Cs233,Canfield_K233}. The crystal of the representing compound \k233 ($T_c=6.1$K) consists of K atom separated (Cr$_3$As$_3$)$^{2-}$ nanotubes (FIG.\ref{fig:geometry}) that are structurally very close to (Mo$_6$Se$_6$)$^{2-}$ in the $M_2$Mo$_6$Se$_6$ (Mo-266) compounds. However, unlike the Mo-266 superconductors, \k233 exhibits peculiar properties at both the normal and the superconducting states. It shows a linear temperature dependent resistivity within a very wide temperature range from 7 to 300K at its normal state; and most interestingly, its upper critical field ($H_{c2}$) steeply increases with respect to decreasing temperature, with extrapolation to 44.7T at 0K, almost four times the Pauli limit. In comparison, the upper critical field of Tl$_2$Mo$_6$Se$_6$ is 5.9T and 0.47T parallel and perpendicular to the quasi-1D direction\cite{PhysRevB.82.235128}, respectively. Such unusually high $H_{c2}$ is beyond the explanation of multiband effect and spin-orbit scattering, and usually implies possible spin triplet pairing. It is therefore quite essential to study the electronic structure of this compound, in particular its ground state, the spin-orbit coupling effect, as well as the possible spin fluctuations.

In this paper, we present our latest first-principles results on the \k233 electronic structure. We show that the ground state of \k233 is paramagnetic. The anti-symmetric spin-orbit coupling (ASOC) effect is negligible far away from the Fermi level, but sizable around it. Combining with the noncentral symmetric nature of the crystal, the sizable ASOC effect around $E_F$ enables spin-triplet pairing mechanism\cite{ncs_jmmm,ncs_book}. The Fermi surfaces of \k233 consist of one 3D sheet and two quasi-1D sheets, and the Cr-3d orbitals, in particular the d$_{z^2}$, d$_{xy}$, and d$_{x^2-y^2}$ orbitals dominate the electronic states around $E_F$. 

\section*{Results}

\begin{table}[ht]
  \caption{Total energies (in meV/Cr) and Cr magnetic moment (in $mu_B$) of considered magnetic states at different optimization levels. The numbers given in/outside the parenthesis for the IAF states are the magnetic moments of Cr$^{\mathrm{I}}$/Cr$^{\mathrm{II}}$ atoms, respectively. The numbers given in and outside the square brackets for the FM states correspond to $m \perp c$-axis and $m\parallel c$-axis, respectively.}
  \begin{tabular}{c||c|c|c|c||c|c||c|c}
    \hline
       & \multicolumn{4}{c||}{GGA} & \multicolumn{2}{c||}{U=1.0 eV} & \multicolumn{2}{c}{U=2.0 eV} \\
    \hline
       & $a$ (\AA) & $c$ (\AA) & $E_{tot}$ & $m_{\mathrm{Cr}}$ & $E_{tot}$ & $m_{\mathrm{Cr}}$  & $E_{tot}$ & $m_{\mathrm{Cr}}$ \\
    \hline
     NM$^{\mathrm{opt}}$  & 10.1030   & 4.1387    &   0  & 0  & \multicolumn{2}{c||}{N/A} & \multicolumn{2}{c}{N/A} \\
     FM$^{\mathrm{opt}}$  & 10.1033   & 4.1399    &  -0.7  & 0.1 & \multicolumn{2}{c||}{N/A} & \multicolumn{2}{c}{N/A} \\
     IAF$^{\mathrm{opt}}$ & \multicolumn{4}{c||}{converge to FM} & \multicolumn{2}{c||}{N/A} & \multicolumn{2}{c}{N/A} \\
    \hline
     NM$^{\mathrm{opt,soc}}$  & 10.1022   & 4.1411    &   0  & 0 & 0 & 0 & 0 & 0\\
     FM$^{\mathrm{opt,soc}}$  & 10.1014   & 4.1412    &  -0.2  & 0.1 & -0.3 & 0.1 & \multicolumn{2}{c}{to IAF}\\
     IAF$^{\mathrm{opt,soc}}$ & \multicolumn{4}{c||}{converge to FM} &  -2.2 & 0.4 (-0.2) & -268.1 & 3.1 (-3.0) \\
     NCL$^{\mathrm{opt,soc}}$ & \multicolumn{4}{c||}{converge to NM} &  -0.8 & 0.3 & -256.8 & 3.1 \\
     IOP$^{\mathrm{opt,soc}}$ & \multicolumn{4}{c||}{converge to NM} &  -6.5 & 1.1 & -330.8 & 2.9 \\
     CLK$^{\mathrm{opt,soc}}$ & \multicolumn{4}{c||}{converge to NM} &  -6.7 & 1.1 & -330.1 & 2.9 \\
    \hline
     NM$^{\mathrm{exp}}$  & \multirow{3}{*}{9.9832}  & \multirow{3}{*}{4.2304} &   0  & 0 & \multicolumn{2}{c||}{N/A} & \multicolumn{2}{c}{N/A} \\
     FM$^{\mathrm{exp}}$  &  &  &  0.0  & 0.2 & \multicolumn{2}{c||}{N/A} & \multicolumn{2}{c}{N/A} \\
     IAF$^{\mathrm{exp}}$ &  &  &  -0.8  & 0.6(-0.5) & \multicolumn{2}{c||}{N/A} & \multicolumn{2}{c}{N/A} \\
    \hline
     NM$^{\mathrm{exp,soc}}$  & \multirow{3}{*}{9.9832}  & \multirow{3}{*}{4.2304} &   0  & 0 & 0 & 0 & 0 & 0\\
     FM$^{\mathrm{exp,soc}}$  &  &  &  0.6 [-0.3]  & 0.2 [0.0] & -2.5 & 0.3 & -9.8 & 0.6 \\
     IAF$^{\mathrm{exp,soc}}$ &  &  & -4.5 & 0.6 (-0.4) & -9.7 & 0.6 (-0.5) & -168.4 & 2.3 (-2.3) \\
     NCL$^{\mathrm{exp,soc}}$ &  &  & \multicolumn{2}{c||}{converge to NM} & -19.7 & 0.7 & -152.7 & 2.3 \\
     IOP$^{\mathrm{exp,soc}}$ &  &  & -6.5 & 0.9 & -46.3 & 1.3 & -273.6 & 2.4 \\
     CLK$^{\mathrm{exp,soc}}$ &  &  & -6.5 & 0.9 & -46.3 & 1.3 & -273.5 & 2.4 \\
    \hline
   \end{tabular}
  \label{tab:state_energy}
\end{table}

We first examine the ground state of \k233. FIG.\ref{fig:geometry} shows the crystal structure of \k233. The Cr atoms form a sub-nanotube along the $c$-axis, surrounded by another sub-nanotube formed by As atoms. The Cr$^{\mathrm{I}}$-Cr$^{\mathrm{I}}$ and Cr$^{\mathrm{II}}$-Cr$^{\mathrm{II}}$ bond lengths are respectively 2.61\AA\ and 2.69\AA, while the Cr$^{\mathrm{I}}$-Cr$^{\mathrm{II}}$ bond length is 2.61\AA. Therefore the six Cr atoms in each unit cell (as well as the six Cr atoms from any two adjacent layers) form a slightly distorted octahedron and become non-centrosymmetric. Since the Cr sub-lattice is magnetically strongly frustruated, we considered several different possible magnetic states in our current study: the ferromagnetic state (FM), the inter-layer antiferromagnetic state (IAF)(FIG. \ref{fig:magiaf}), the noncollinear antiferromagnetic state (NCL)(FIG. \ref{fig:magncl}), the noncollinear in-out state (IOP) (FIG. \ref{fig:magiop}), the noncollinear chiral-like state (CLK) (FIG. \ref{fig:magclk}), and the nonmagnetic state (NM). The total energy calculation results are shown in TAB. \ref{tab:state_energy}. Without structural optimization, all the magnetically stable states (FM, IAF, IOP and CLK) are energetically degenerate within the DFT errorbar. The presence of multiple nearly degenerate magnetic configurations indicates that the ground state is indeed paramagnetic phase. It is important to notice that the converged IAF state yields -0.40 $\mu_B$ and 0.58 $\mu_B$ magnetic moment for Cr$^{\mathrm{I}}$ and Cr$^{\mathrm{II}}$, respectively;  although it is initially constructed with zero total magnetization per unit cell. Such difference leads to a non-zero total magnetization in one unit cell, therefore the IAF state is indeed a ferrimagnetic state. This phase cannot be stabilized in DFT unless the effective inter-cell magnetic interaction is ferromagnetic. With structural optimization, the IAF state will eventually converge to FM state which is also energetically degenerate with NM state within the DFT errorbar. The above facts suggest that the system is close to magnetic instabilities but long range order cannot be established. Therefore, we argue that the ground state of \k233 is paramagnetic (PM) state. 

We have also performed GGA+$U$ calculations with $U$=1.0 and 2.0 eV, $J$=0.7 eV for the compound. The inclusion of on-site $U$ strengthened the magnetic interactions, leading to more stable magnetic states. As a result, all the non-collinear phases became energetically stable upon structural optimization even with a small $U$=1.0 eV. Nevertheless, these states are still energetically degenerate within LDA errorbar, leaving the above conclusion unaltered. However, the staggered moment per Cr atom become $\sim$ 30 times the experimental value with $U$=2.0 eV. Thus the GGA+$U$ method significantly overestimates the magnetism in this system, suggesting that the GGA functional better suits the current study.

With closer examination of TAB. \ref{tab:state_energy}, it is important to notice that after full structural optimization (where both lattice constants and atomic coordinates are allowed to relax), the resulting $a$ is 1.2\% larger than the experimental value, but $c$ is 2\% shorter. Similar situation was also observed in the studies of iron-based superconductors\cite{singh_1111}, and was due to the fact that the NM state in LDA is only a very rough approximation to the actual PM state without considering any magnetic fluctuations. As the electronic structure close to $E_F$ seems to be sensitive to the structural changes, all the rest reported results are based on the experimental geometry unless specified, following the examples of iron-pnictides/chalcogenides.

  \begin{figure}
    \subfigure[]{
      \rotatebox{270}{\includegraphics[width=6cm]{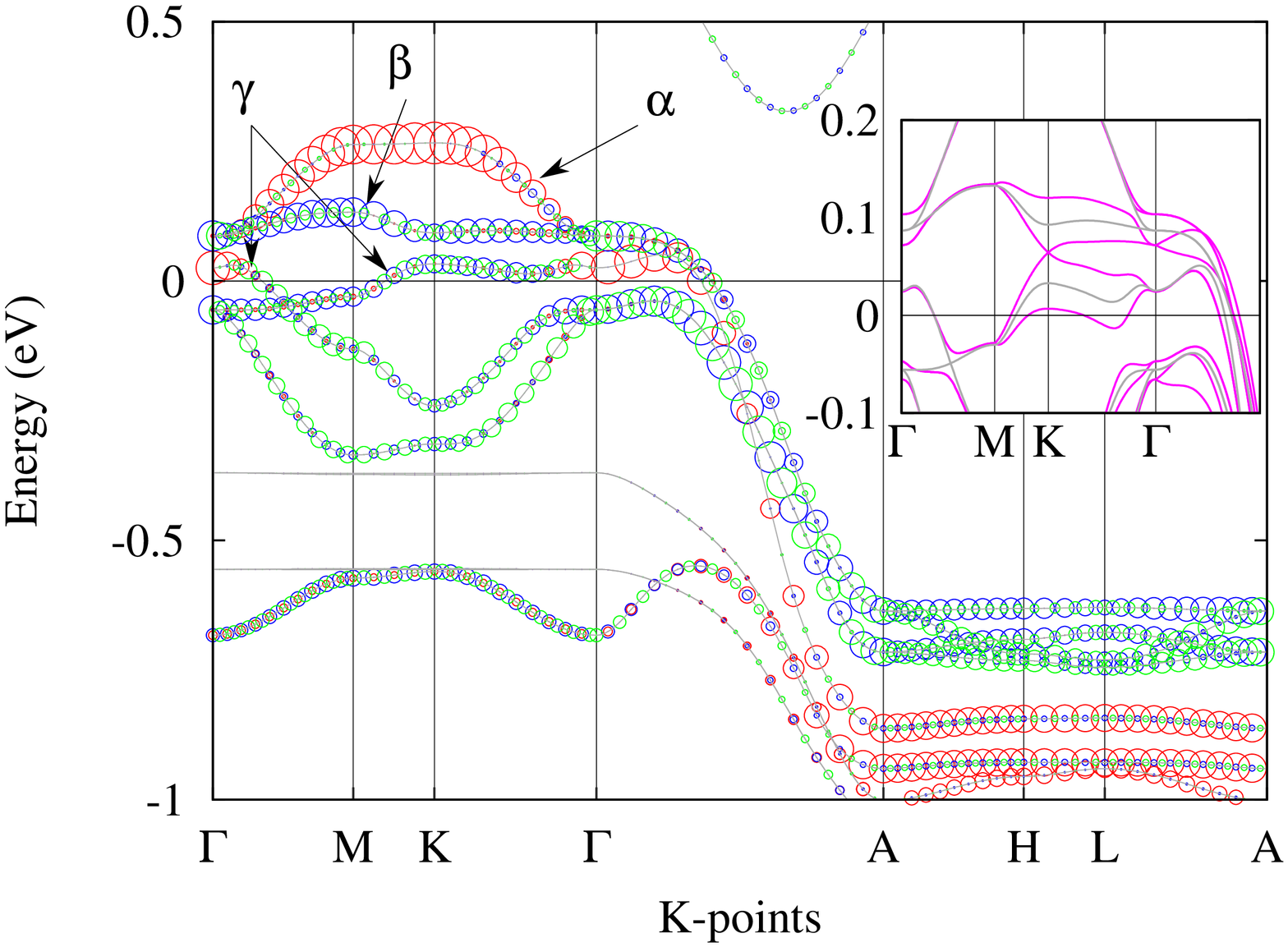}}
      \label{fig:bs}}
    \subfigure[]{
      \rotatebox{270}{\includegraphics[width=6cm]{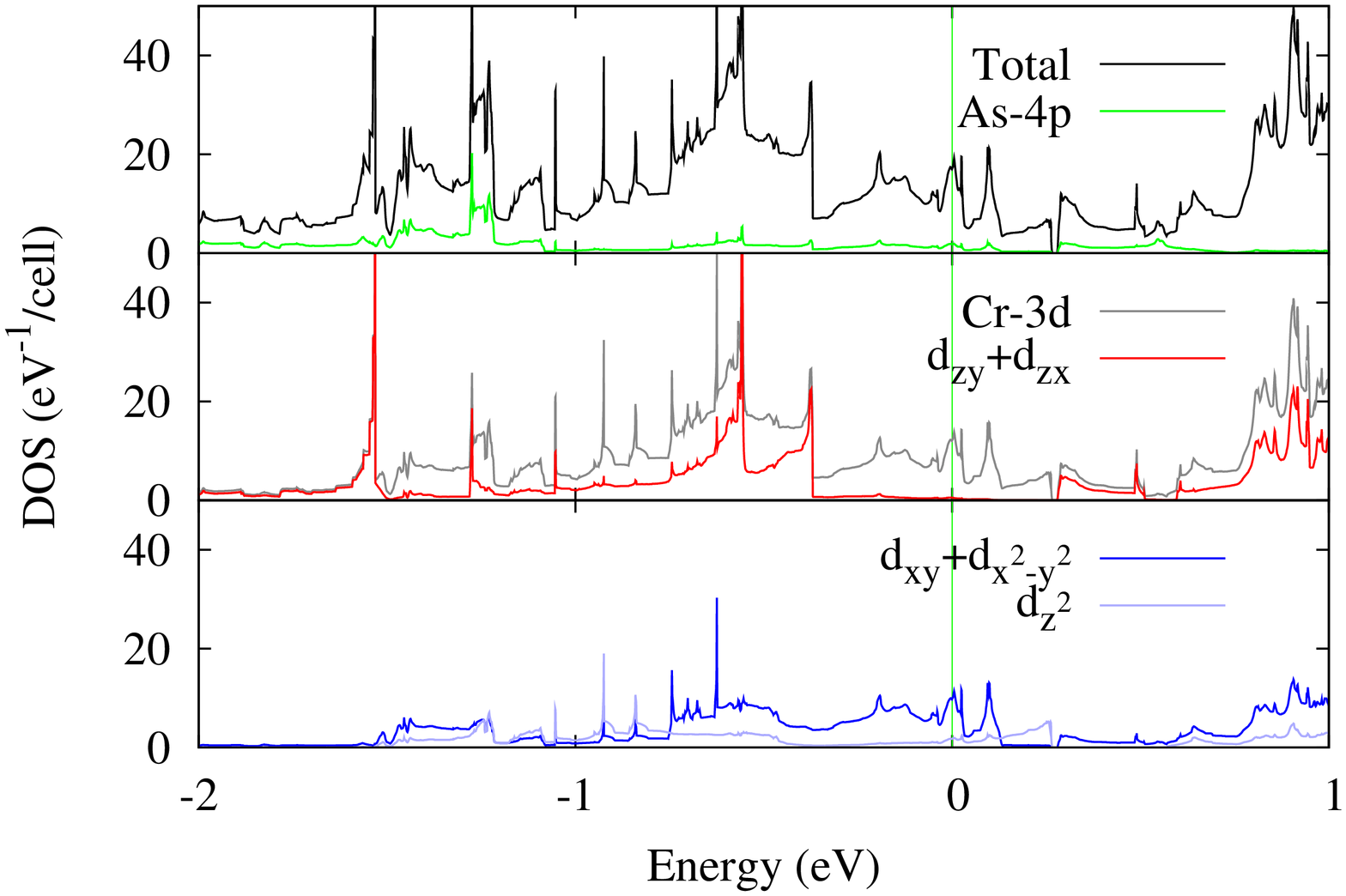}}
      \label{fig:dos}}
    \caption{(a)Electronic band structure of \k233. The inset shows comparison between relativistic (magenta solid lines) and non-relativistic (grey solid lines) results. The size of the red, blue and green circles are proportional to the contributions from the d$_{z^2}$, d$_{x^2-y^2}$ and d$_{xy}$ orbitals, respectively. (b) Electronic density of state of \k233 from the non-relativistic calculation. Upper panel shows the total DOS (black line) and PDOS from As-4p (green line); middle and bottom panel shows the PDOS from all five-orbitals of Cr-3d (grey lines in the middle panel), the PDOS of Cr-3d$_{zx}$ and Cr-3d$_{zy}$ (red line in the middle panel), and the PDOS of Cr-3d$_{z^2}$ (light blue line in the bottom panel), Cr-3d$_{x^2-y^2}$ and Cr-3d$_{xy}$ (blue line in the bottom panel). \label{fig:bsdos}}
  \end{figure}

Figure \ref{fig:bsdos} shows the band structure and density of states (DOS) of \k233. The band structure (FIG. \ref{fig:bs}) shows flat bands at $k_z=0.0$ and $k_z=0.5$ around -0.93 eV, -0.86 eV, -0.64 eV, -0.56 eV, -0.37 eV, 0.51 eV and 0.95 eV relative to $E_F$, forming quasi-1D van Hove singularities in the DOS plot at corresponding energies. All bands including the flat bands mentioned above disperse significantly along $k_z$, which clearly indicates the quasi-1D nature of the compound. From the band structure plot, it is apparent that there are three bands cross the Fermi level, which are denoted as $\alpha$, $\beta$ and $\gamma$ bands as shown in FIG. \ref{fig:bs}, respectively. The $\alpha$ and $\beta$ bands are degenerate along $\Gamma$-A in the non-relativistic calculations, and they do not cross the Fermi level in  either the $k_z=0$ plane nor the $k_z=0.5$ plane, thus are likely to be quasi-1D bands. Instead, the $\gamma$ band crosses the Fermi level not only along $\Gamma$-A, but also in the $k_z=0$ plane, therefore it is likely to be a 3D band. The total DOS at $E_F$ is calculated to be $N(E_F)$=8.58 states/(eV$\cdot$f.u.) under non-relativistic calculations, with approximately 8\%, 17\% and 75\% contribution from $\alpha$, $\beta$ and $\gamma$ bands, respectively. Experimentally, the measured specific heat data yields $\gamma$=70 mJ/(K$^2\cdot$mol), which corresponds to $N(E_F)\approx$29.7 states/(eV$\cdot$f.u.), suggesting an electron mass renormalization factor $\approx3.5$, comparable to the one in the iron-pnictide case. The projected density of states (PDOS) (FIG. \ref{fig:dos}) shows that the Cr-3d orbitals dominate the electronic states from $E_F$-1.0 eV to $E_F$+0.5 eV, and more importantly, the states from $E_F$-0.36 eV to $E_F$+0.26 eV are almost exclusively from the d$_{z^2}$, d$_{xy}$, and d$_{x^2-y^2}$ orbitals. It is also instructive to notice that the five bands near $E_F$ has a band width of only $\approx$0.6 eV at the $k_z=0$ plane in our current calculation, and would be $\approx$0.2 eV if the electron mass renormalization factor is considered. Thus these bands are extremely narrow at $k_z=0$ plane, consistent with the quasi-1D nature of the compound.

We have also performed relativistic calculations to identify the spin-orbit coupling (SOC) effect. Since the magnitude of SOC effect is proportional to the square root of the atomic number, the SOC effect is initially expected to be negligible for this compound. Indeed, the SOC effect on the $\alpha$ band is negligible. However, it is notable that the SOC effect lifted the degeneracy between $\alpha$ and $\beta$ bands at $\Gamma$, as well as along $\Gamma$-A. More importantly, sizable ($\approx$60 meV) ASOC splitting can be identified on $\beta$ and $\gamma$ bands around K close to $E_F$(insets of FIG. \ref{fig:bs}). Considering that the crystal structure lacks of an inversion center, the presence of ASOC splitting facilitates the spin-triplet pairing, which is a natural and reasonable explanation of its unusually high $H_{c2}$ \cite{ncs_jmmm,ncs_book}. 

  \begin{figure}
    \subfigure[3D view]{
      \includegraphics[width=2cm]{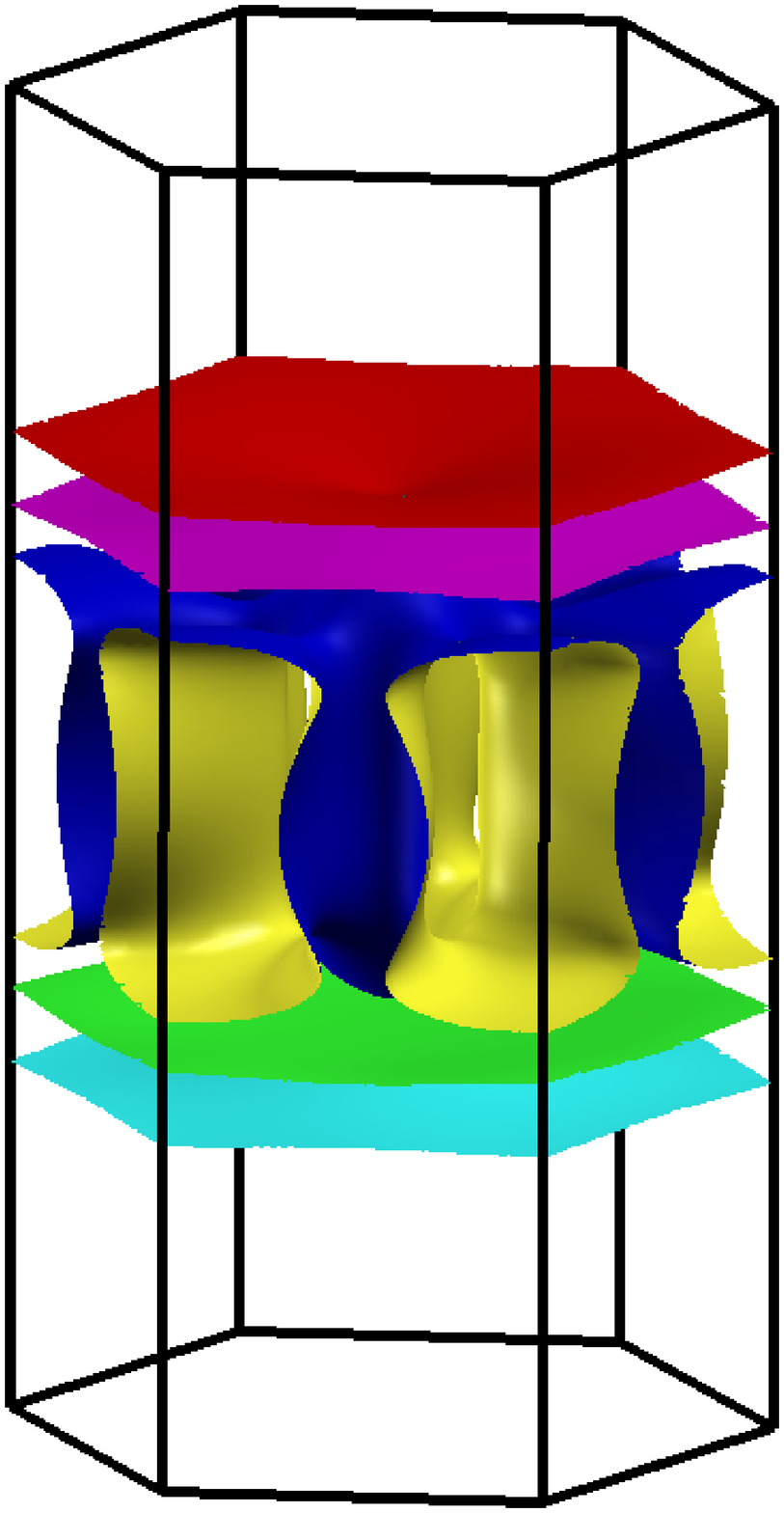}
    }
    \subfigure[Side view]{
      \includegraphics[width=2cm]{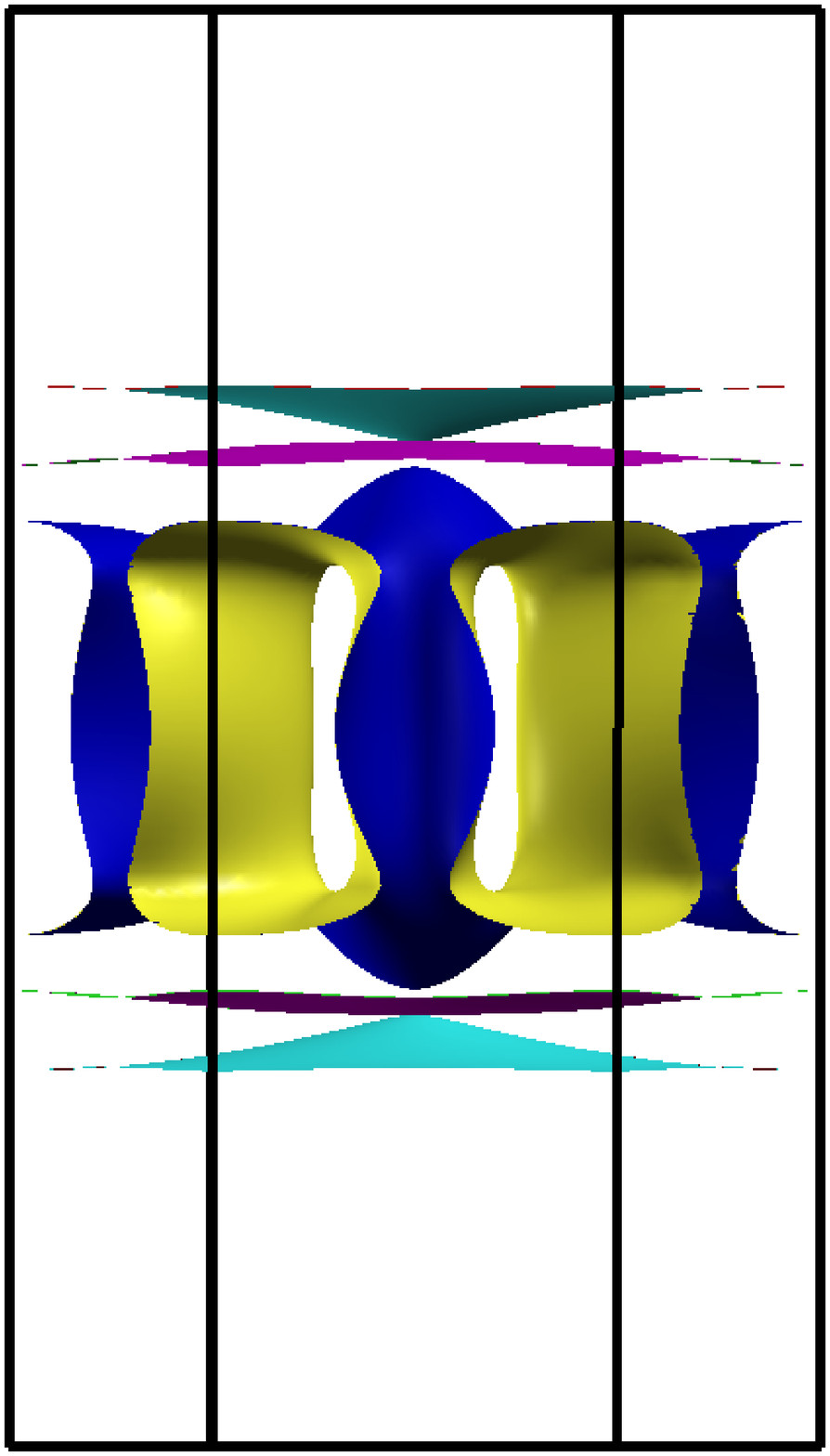}
    }
    \subfigure[$\alpha$ band]{
      \includegraphics[width=2cm]{fig3c}
    }
    \subfigure[$\beta$ band]{
      \includegraphics[width=2cm]{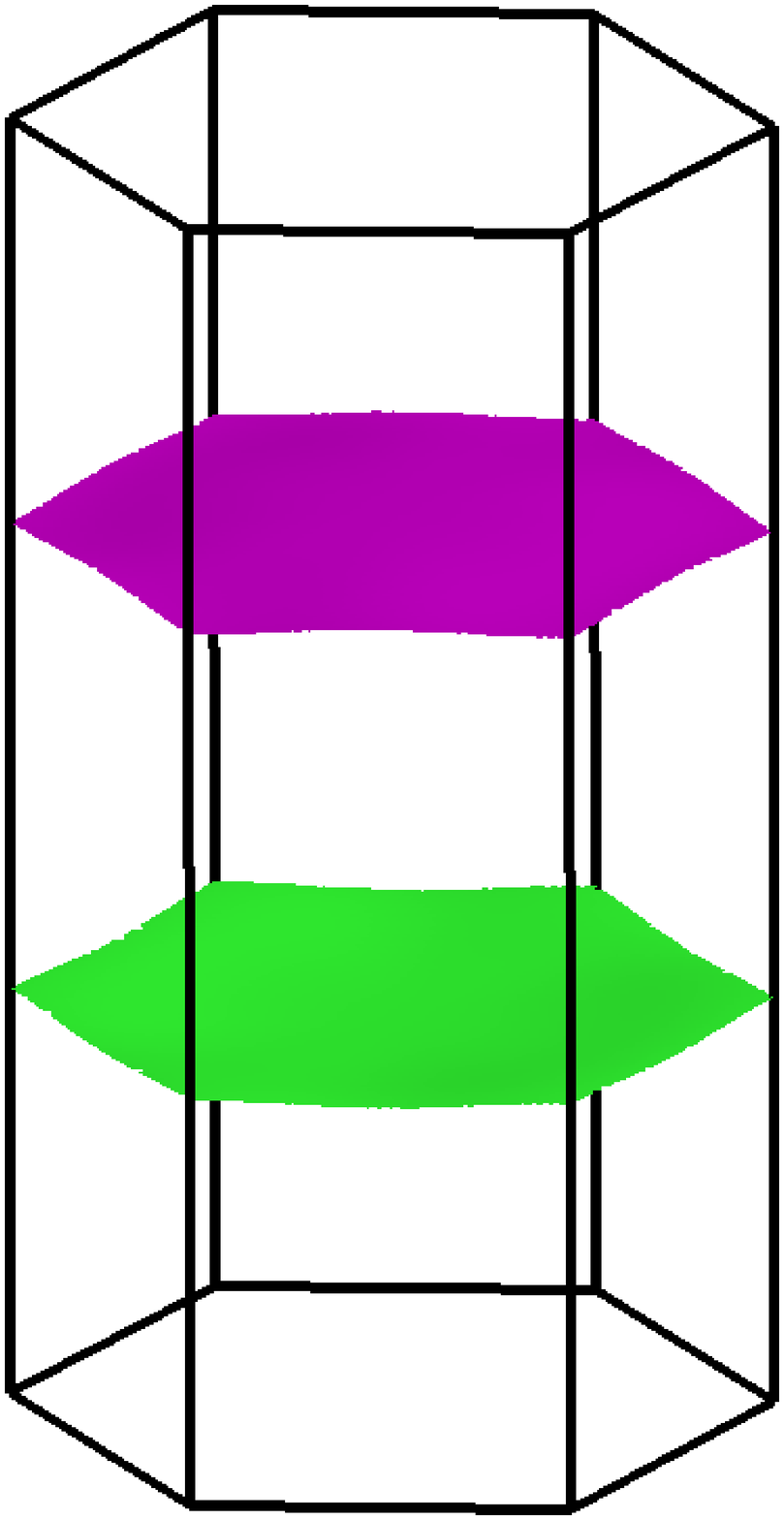}
    }
    \subfigure[$\gamma$ band]{
      \includegraphics[width=2cm]{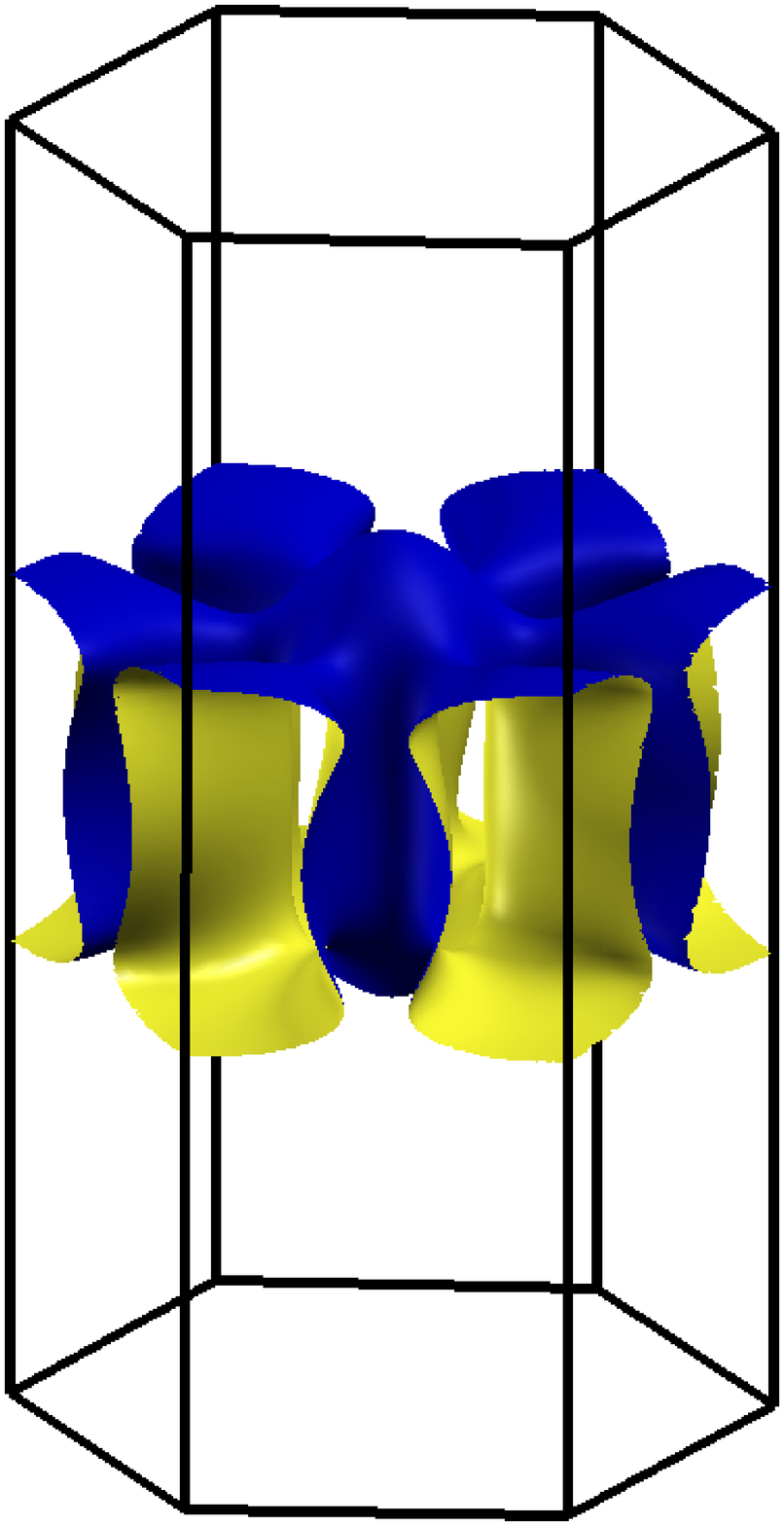}
    }
    \subfigure[$\gamma$ band (Top view)]{
      \includegraphics[width=3cm]{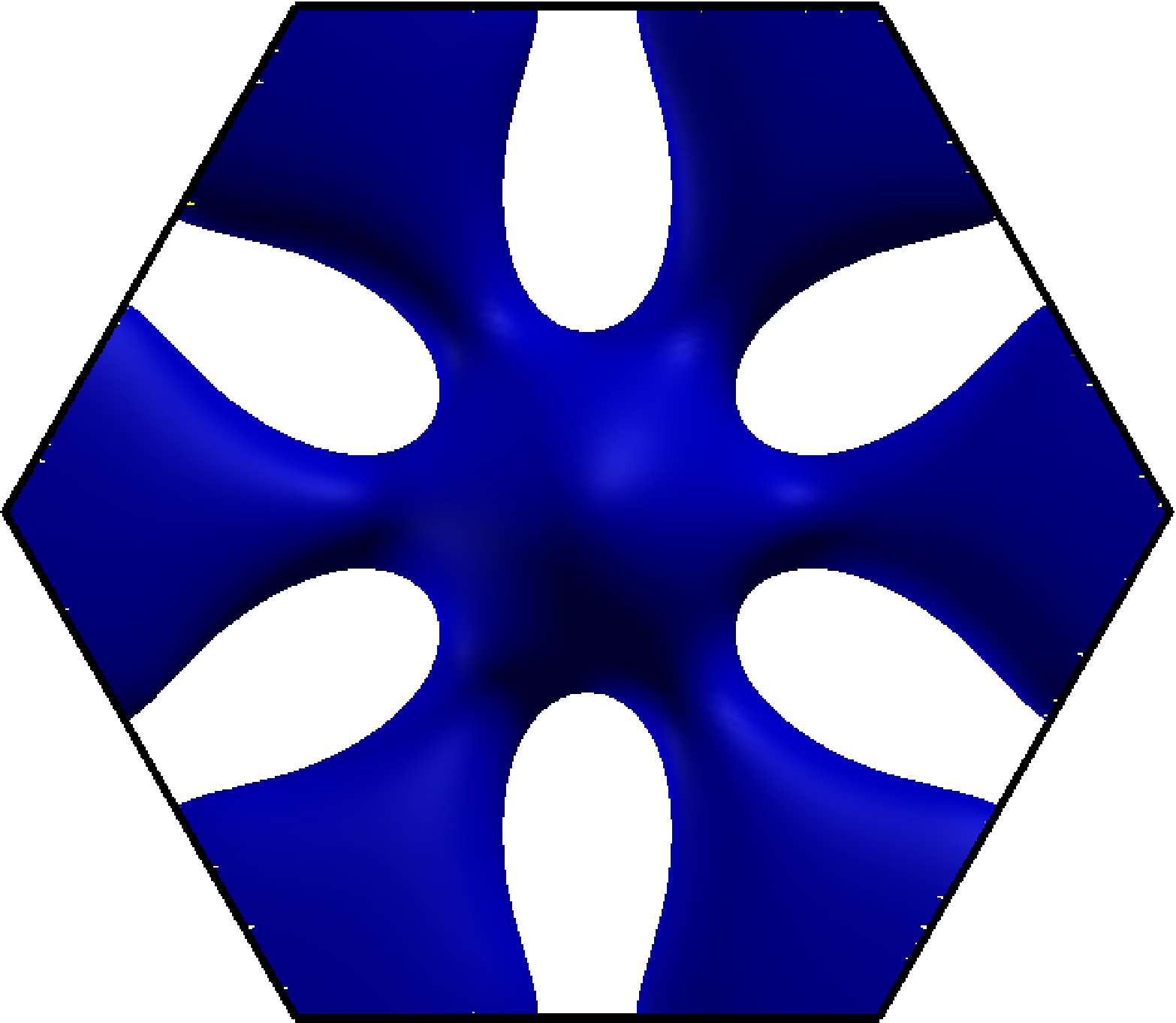}
    }
    \subfigure[$\gamma$ band, SOC]{
      \includegraphics[width=2cm]{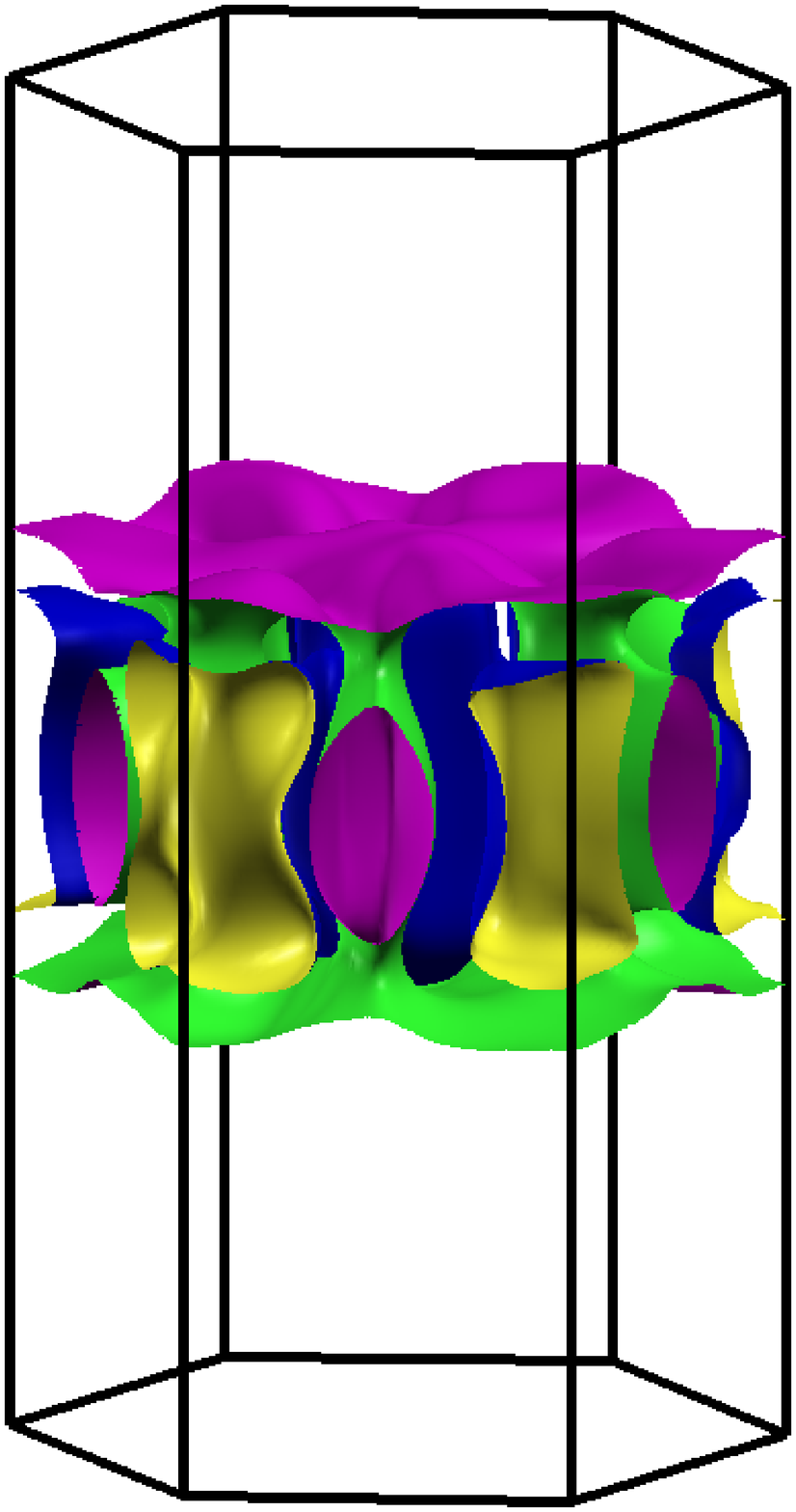}
    }
    \caption{Fermi surface sheets of \k233. (a-f) Non-relativistic results (SOC effect not included). (g) Relativistic result for $\gamma$ band (with SOC effect). The irreducibie Brillouin zone and high symmetry points are indicated in (c). In all figures, $\Gamma$ point is at the zone center. \label{fig:fs}}
  \end{figure}

We plot the Fermi surface sheets of \k233 in FIG. \ref{fig:fs}. The non-relativistic Fermi surface of \k233 consists of one 3D sheet formed by the $\gamma$ band and two quasi-1D sheets formed by the $\alpha$ and $\beta$ bands. These sheets cut $k_z$ axis at $k_F^{\alpha}$=$k_F^{\beta}$=0.30 \AA$^{-1}$ and $k_F^{\gamma}$=0.27 \AA$^{-1}$, respectively. Should the SOC effect taken into consideration, the $\gamma$ band is mostly affected and its Fermi surface sheet splits into two (FIG. \ref{fig:fs}g). Nevertheless the two splitted $\gamma$ sheets cut $k_z$ axis at the same point at $k_F^{\gamma}$=0.25 \AA$^{-1}$. For the $\beta$ band, the SOC effect causes the splitting of its Fermi surface sheet, forming two adjacent quasi-1D sheets who cut $k_z$ axis at $k_F^{\beta,I}$=0.31 \AA$^{-1}$ and $k_F^{\beta,II}$=0.33 \AA$^{-1}$, respectively. The SOC effect on $\alpha$ band is negligible, thus the $\alpha$ band and its Fermi surface remain mostly unaltered and $k_F^{\alpha}$ remains unchanged.

  \begin{figure}
    \rotatebox{270}{\includegraphics[width=6cm]{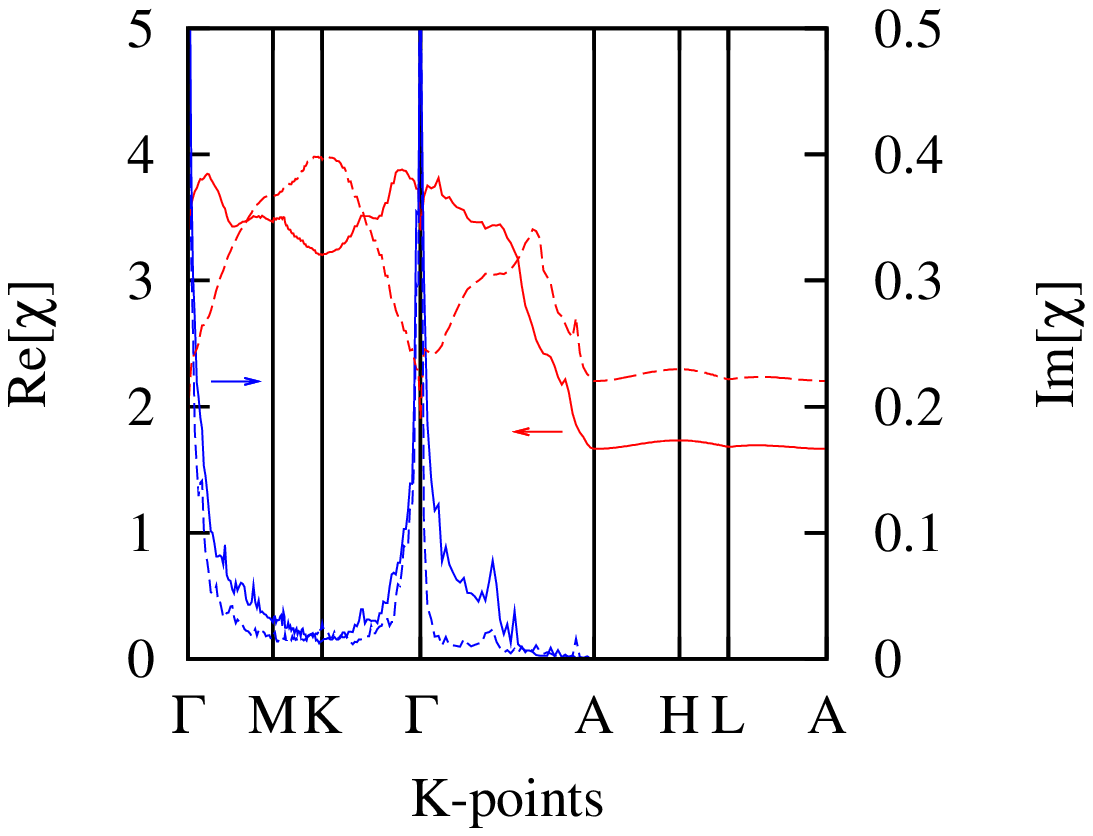}}
    \rotatebox{270}{\includegraphics[width=6cm]{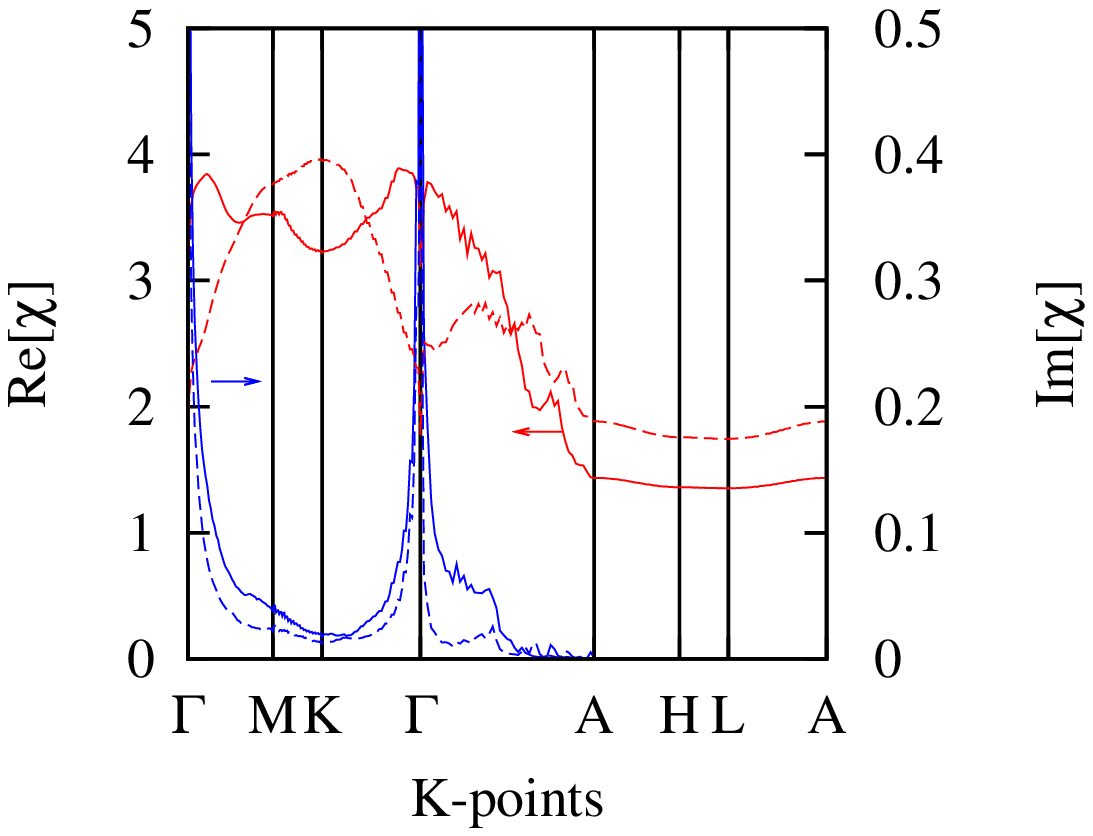}}
    \caption{Static bare electron susceptibility $\chi_0(\omega=0)$ of \k233 (solid lines) and 0.2 hole doped \k233 (K$_{1.8}$Cr$_3$As$_3$) (dashed lines) calculated using (a) 5 molecular orbitals localized on Cr-octahedron and (b) 10 molecular orbitals localized on Cr-octahedron. The red lines denote the real part $\mathrm{Re}[\chi_0(\omega=0)]$; and the blue lines are the imaginary part $\lim_{\omega\rightarrow0}\mathrm{Im[\chi_0(\omega, \mathbf{q})]/\omega}$. \label{fig:chi}}
  \end{figure}

We have also calculated the bare electron susceptibility $\chi_0$ (FIG. \ref{fig:chi}). The real part of the susceptibility shows no prominent peak, consistent with our PM ground state result. However, the imaginary part of the susceptibility shows extremely strong peaks at $\Gamma$, which may be due to the large electron DOS at the Fermi level $E_F$. Such a strong peak of imaginary part of susceptibility at $\Gamma$ is usually an indication of large FM spin fluctuation when finite on-site repulsion is taken into consideration. Considering the fairly large ASOC effect, together with the non-central symmetric nature of the crystal structure, the large FM spin fluctuation may be the pairing mechanism and contribute to the spin-triplet pairing channel. It should be noted that the strong peak of susceptibility imaginary part at $\Gamma$ is robust with respect to hole doping, as our rigid-band calculation of 0.2 hole doping shows the same feature.

\section*{Discussion}

Our above studies are focused on calculations with experimental structure parameters. However, it is important to point out that the number of Fermi surface sheets and their respective dimensionality characters, the orbital characters of the bands close to $E_F$, the SOC effect on these bands, and the features of $\chi_0$ are insensitive with respect to the structural optimization; therefore the physics is robust against the structural change (Supplementary Information, FIG. SI-2, SI-4 and SI-6). Nevertheless, the detailed band structure of \k233 is quite sensitive to the structure relaxation. In particular, the 3D $\gamma$ band will submerge below $E_F$ at $\Gamma$ and the 3D $\gamma$ sheet will be topologically different from the one shown in FIG. \ref{fig:fs} (Supplementary Information, FIG. SI-3 and SI-5). The quasi-1D $\alpha$ and $\beta$ bands are much less affected, although they intersect with the $k_z$ axis at different $k_F$s. 

In conclusion, we have performed first-principles calculations on the quasi-1D superconductor \k233. The ground state of \k233 is PM. The electron states near the Fermi level are dominated by the Cr-3d orbitals, in particular the 3d$_{z^2}$, d$_{xy}$, and d$_{x^2-y^2}$ orbitals from $E_F$-0.5 eV to $E_F$+0.5 eV. The electron DOS at $E_F$ is less than 1/3 of the experimental value, suggesting intermediate electron correlation effect may in place. Three bands cross $E_F$ to form two quasi-1D sheets in consistent with its quasi-1D feature, as well as one 3D sheet who is strongly affected by the ASOC splitting as large as 60 meV. Finally, the real part of the susceptibility is mostly featureless, consistent with our PM ground state result, but the imaginary part of the susceptibility shows large peaks at $\Gamma$, indicating large ferromagnetic spin fluctuation exists in the compound. Combined with the large ASOC effect and the lack of inversion center in the crystal structure, the experimentally observed abnormally large $H_{c2}$(0) may be due to a spin-triplet pairing mechanism.

\section*{Methods}

First-principles calculations were done using VASP\cite{vasp_1} with projected augmented wave method\cite{bloch_paw,vasp_2} and plane-wave. Perdew, Burke and Enzerhoff flavor of exchange-correlation functional\cite{PBE_xc} was employed. The energy cut-off was chosen to be 540 eV, and a 3$\times$3$\times$9 $\Gamma$-centered K-mesh was found sufficient to converge the calculations. The K-3p, Cr-3p and As-3d electrons are considered to be semi-core electrons in all calculations.

The Fermi surfaces were plotted by interpolating a five orbital tight-binding Hamiltonian fitted to LDA results using the maximally localized Wannier function (MLWF) method\cite{PhysRevB.56.12847,Mostofi2008685}. The same Hamiltonian were also employed in the calculation of the bare electron susceptibility $\chi_0$ using:
\begin{align*}
\chi_0(\omega, \mathbf{q})=\frac{1}{N_{\mathbf{k}}}\sum_{mn}\sum_{\mu\nu\mathbf{k}}\frac{\langle m\vert\mu\mathbf{k}\rangle\langle\mu\mathbf{k}\vert n\rangle\langle n\vert\nu\mathbf{k+q}\rangle\langle\nu\mathbf{k+q}\vert m\rangle}{\omega+\epsilon_{\nu\mathbf{k+q}}-\epsilon_{\mu\mathbf{k}}+i0^{+}} \times\left[f(\epsilon_{\mu\mathbf{k}})-f(\epsilon_{\nu\mathbf{k+q}})\right]
\end{align*}
where $\epsilon_{\mu\mathbf{k}}$ and $f(\epsilon_{\mu\mathbf{k}})$ are the band energy (measured from $E_F$) and occupation number of $\vert \mu\mathbf{k}\rangle$, respectively; $\vert n\rangle$ denotes the $n^{th}$ Wannier orbital; and $N_{\mathbf{k}}$ is the number of the $\mathbf{k}$ points used for the irreducible Brillouin zone (IBZ) integration. The value of $\chi_0$ at $\mathbf{q}=\Gamma$ was approximated by the value at $\mathbf{q}=(\delta, 0, 0)$ with $\delta=0.001$.


\section*{Acknowledgements}

The authors would like to thank Fuchun Zhang, Zhuan Xu, Yi Zhou and Jianhui Dai for inspiring discussions. This work has been supported by the NSFC (No. 11274006 and No. 11190023), National Basic Research Program (No. 2014CB648400 and No. 2011CBA00103) and the NSF of Zhejiang Province (No. LR12A04003). All calculations were performed at the High Performance Computing Center of Hangzhou Normal University College of Science.

\section*{Author contributions statement}

H. J. performed most of the calculations; G.-H. C. and C. C. were responsible for the data analysis and interpretation; C. C. calculated the response functions and drafted the manuscript.

\section*{Additional information}

\textbf{Competing financial interests} The authors declare no competing financial interests.

\end{document}